\documentclass[12pt]{article}
\usepackage{epsfig}
\def\figcap{\section*{Figure Captions\markboth
        {FIGURECAPTIONS}{FIGURECAPTIONS}}\list
        {Figure \arabic{enumi}:\hfill}{\settowidth\labelwidth{Figure 999:}
        \leftmargin\labelwidth
        \advance\leftmargin\labelsep\usecounter{enumi}}}
 \relax
\def\deg{^\circ}
\def\e+e-{{$e^+e^-$ }}
\def\ep{$ep$ }
\def\Leff{\Lambda_{\hbox{\it eff}}}
\def\avn{$\left< n \right>$\ }
\def\mavn{\left< n \right>}
\def\mQ{$\left< Q \right>$\ }
\def\CPC#1#2#3{Computer Phys. Comm. #1 (19#3) 32}
\def\PL#1#2#3{Phys.\ Lett.\ #1B (19#3) #2}
\def\PR#1#2#3{Phys.\ Rev.\ #1D (19#3) #2}
\def\NP#1#2#3{Nucl.\ Phys.\ #1B (19#3) #2}
\def\NC#1#2#3{Nuovo Cimento\ #1A (19#3) #2}
\def\ZP#1#2#3{Z.\ Phys.\ C#1 (19#3) #2}

\def\NIM#1#2#3{Nucl.\ Instr.\ Meth. A#1 (19#3) #2}

\begin{document}
\begin{flushleft}
  {\tt DESY 97-108}\hfill {\tt ISSN 0418-9833} \\
  {\tt June 1997} 
\end{flushleft}
\vspace*{1.8cm}
\begin{center}
\begin{Large}
\boldmath
{\bf     Evolution of \ep Fragmentation and Multiplicity     \\
         Distributions in the Breit Frame
\\}
\unboldmath 
\end{Large}
\vspace*{1cm}
\begin{Large}
H1 Collaboration \\
\vspace*{0.5cm}
\end{Large}

\end{center}

\begin{abstract}
\noindent 
        Low $x$ deep-inelastic \ep scattering data, taken in 1994 at the H1 
detector at HERA, are analysed in the Breit frame of reference. 
The evolution of the peak and width of the current hemisphere fragmentation 
function is presented as a function of $Q$ and compared 
with \e+e- results at equivalent centre of mass energies.
Differences between the average charged multiplicity and the multiplicity
of \e+e- annihilations at low energies are analysed.
Invariant energy spectra are compared with MLLA predictions.
Distributions of multiplicity are presented as functions of Bjorken-$x$
and $Q^2$, and KNO scaling is discussed.

\end{abstract}
\vfill
%\hrule
%\begin{center}
%\tt
%Editors:\qquad
%P.~Dixon, D.~Kant, G.~Thompson
%\\
%Referees:\qquad  K.~Mueller, J.~Turnau\\
%Final comments to the above by 13:00 Thursday 12th May, 1997.  \\
%\rm
%\end {center}
% Comment the following 3 lines for no page numbers
%\clearpage\mbox{}\clearpage

\setcounter{page}{1}
\pagestyle{plain}

\newpage
\noindent
%\input qmphvc$dka300:[workspace0.thompson.h1dat]h1auts.tex
%   H1AUTS  Author list by names, no. of authors  392
%           status: 24/02/97   09.15.43
 C.~Adloff$^{35}$,                %WUPP-ST                  Adloff              
 S.~Aid$^{13}$,                   %HAM2-LEFT    8/96        Aid                 
 M.~Anderson$^{23}$,              %MANC-ST  10/95           Anderson            
 V.~Andreev$^{26}$,               %LPI -PD                  Andreev             
 B.~Andrieu$^{29}$,               %ECPL-PD                  Andrieu             
 V.~Arkadov$^{36}$,               %ZEUT-ST    10/96         Arkadov             
 C.~Arndt$^{11}$,                 %DESY-ST   1/96           Arndt               
 I.~Ayyaz$^{30}$,                 %PARI-ST       5/96       Ayyaz               
 A.~Babaev$^{25}$,                %ITEP-PD                  Babaev              
 J.~B\"ahr$^{36}$
% ...
%ZUER-ST                  ZurNedden           

\vskip 3mm
\noindent
%\input qmphvc$dka300:[workspace0.thompson.h1dat]h1inst.tex
%     H1 Institutes as appearing on publications
% ...
$ ^{11}$ DESY, Hamburg, Germany$^a$ \\
% ...
$ ^{13}$ II. Institut f\"ur Experimentalphysik, Universit\"at Hamburg,
                                     Hamburg, Germany$^ a$  \\
% ...
$ ^{23}$ Physics Department, University of Manchester,
                                          Manchester, UK$^ b$\\
% ...
$ ^{25}$ Institute for Theoretical and Experimental Physics,
                                                 Moscow, Russia \\
 $ ^{26}$ Lebedev Physical Institute, Moscow, Russia$^ f$ \\
% ...
$ ^{29}$ LPNHE, Ecole Polytechnique, IN2P3-CNRS,
                             Palaiseau, France \\
 $ ^{30}$ LPNHE, Universit\'{e}s Paris VI and VII, IN2P3-CNRS,
                              Paris, France \\
% ...
$ ^{35}$ Fachbereich Physik, Bergische Universit\"at Gesamthochschule
               Wuppertal, Wuppertal, Germany$^ a$ \\
 $ ^{36}$ DESY, Institut f\"ur Hochenergiephysik,
                              Zeuthen, Germany$^ a$\\
% ...
$ ^a$ Supported by the Bundesministerium f\"ur Bildung, Wissenschaft,
        Forschung und Technologie, FRG,
        under contract numbers 6AC17P, 6AC47P, 6DO57I, 6HH17P, 6HH27I,
        6HD17I, 6HD27I, 6KI17P, 6MP17I, and 6WT87P \\
 $ ^b$ Supported by the UK Particle Physics and Astronomy Research
       Council, and formerly by the UK Science and Engineering Research
       Council \\
% ...
$ ^f$ Supported by the Deutsche Forschungsgemeinschaft \\
% ...

\newpage
\section{Introduction}

In a previous publication~\cite{93BF} the H1 experiment analysed hadron 
production in \ep deep--inelastic scattering (DIS) in the 
Breit frame~\cite{Jones,Streng} of reference. We showed similarities between 
hadronic distributions in the `current' hemisphere of this frame of reference 
and those in a given single hemisphere of an \e+e- annihilation interaction. 
A similar analysis~\cite{ZEUS} was performed by the ZEUS experiment. The
first H1 study analysed 1993 data resulting from $e^-p$ collisions. The 1994 
run used incident positrons and this analysis has more than six times as many 
events despite tighter cuts on event quality. Since the $Q^2$ values 
investigated are still in general below the kinematic region where anything 
other than virtual photon exchange is relevant, the results of the two 
analyses are directly comparable.

        It is the object of this analysis to study the evolution of the
fragmentation function and charged hadronic multiplicity distributions. 
One aim is to continue the comparison with equivalent \e+e- results as a test 
of the universality of properties of the partons found within a proton and 
those of the quarks produced in pairs from the vacuum. Another aim is to show
sensitivity to the running of the QCD coupling constant and to leading order 
(LO) QCD processes from consideration of such hadronic final states.

        After a description of the relevant parts of the apparatus (section 2)
and a discussion of the treatment of the data (section 3), this paper describes
the Breit frame of reference (section 4) and extends (section 5) the earlier 
analysis of the evolution of the fragmentation function with $Q^2$, where 
$q^2 = -Q^2$ is the square of the four-momentum transferred from the incident 
lepton. The average charged multiplicity dependence on $Q$ (section 6) is
used to study the effects of LO QCD processes. For the first time we present  
invariant energy spectra (section 7) from HERA and H1 multiplicity 
distributions (section 8) in the Breit frame.

\section{The H1 Detector} 

         The H1 detector is described in detail elsewhere~\cite{H1app}. Here, 
we give only a short description of the components more relevant to this
analysis. We employ a right-handed coordinate system with the
positive (forward) $z$ axis being in the direction of the incoming proton
beam. Polar angles are measured from this direction.

        Momentum measurements of charged particles are provided, in the central 
region of the apparatus, by two cylindrical and co-axial drift 
chambers~\cite{H1CDT} for ($r,\phi$) measurement which have significantly fewer
dead cells compared with the situation in 1993. These detectors are 
supplemented by two $z$-chambers. In the forward (proton) direction the 
equivalent measurements are provided by three Radial and three Planar drift 
chamber modules, the Forward Track Detector~\cite{FTD}. All these track 
detectors are inside a uniform 1.15~T magnetic field. Track segments from all 
devices are combined to give efficient detection and momentum measurement with 
$\delta p/p^2$ {\raisebox{-0.7ex}{$\stackrel{\textstyle<}{\sim}$}}1\% /GeV for 
most of the angular range used in this analysis, $10\deg <\theta<160\deg $. 

        In the polar angle range $4\deg <\theta<153\deg $ the trackers 
are surrounded by a fine-grained liquid argon (LAr) sampling 
calorimeter~\cite{H1CAL} with lead and steel absorber in the electromagnetic 
and hadronic sections respectively. The calorimeter cells measure hadronic 
energy flow and the energy of the scattered electron for high $Q^2$ events. 
The LAr calorimeter is complemented by a backward electromagnetic, 
lead-scintillator, calorimeter (BEMC~\cite{BEMC}) covering the angular range 
$151\deg <\theta<176\deg $. Behind this there is a double--layer scintillator
hodoscope which gives efficient background rejection by time of flight
measurement. The data are derived from two separate samples corresponding to 
the scattered positron being triggered by, and detected in, either the BEMC or 
the LAr calorimeter. The transition is at a $Q^2$ of about 100~GeV$^2$ and these
samples are referred to in this paper as, respectively, low or high $Q^2$ data.
The trigger is already fully efficient at positron energies of 10 GeV.
 
\section{Data Selection and Corrections} 

        The full data sample, triggered as above, consists of some $150\,000$ 
events taken at an \ep centre of mass energy, $\sqrt s$, of 300 GeV. Event 
kinematic variables used in this analysis are calculated using only the 
scattered lepton, which gives both the best resolution in the chosen region 
and clear freedom from bias on the hadronic system studied. Throughout, the 
events are selected to have an identified scattered positron~\cite{F294} with 
an energy $E_e^{\prime}>14$~GeV.

        To reject beam associated background, we demand that there be no veto
from the time of flight system and that a vertex is found for the event 
within 30~cm of the nominal vertex position. The low (high) $Q^2$ data 
referred to in the above section are selected to be within the limits  
$12<Q^2<100$~GeV$^2$ ($100<Q^2<8000$~GeV$^2$). The dimensionless inelasticity 
variable, $y= Q^2/xs$,  is required to be in the range  $0.05<y<0.6$. These 
conditions ensure that contamination from mis-identified photoproduction events
 is below the 1\% level. Together these cuts remove $\sim$ 45\% of the original
sample. In addition, a cut is placed to safely exclude a further $\sim$ 5\%
of diffractive events which are not well-modelled with the DIS Monte Carlo 
programs used in this analysis and which have no analogue in \e+e- 
interactions. We require that a total cluster energy of at least 0.5~GeV 
should be observed in the region $4.4\deg <\theta < 15\deg$.

        Events are then selected in which a massless quark would be scattered 
through $10\deg <\theta <150\deg$ using four momentum conservation with the 
scattered positron.  This is inside the acceptance of the H1 track detectors 
and thus minimises corrections, at the cost of removing a further $\sim$ 24\% 
of events. 

There is a source of error arising from QED radiation which comes
about because of an incorrect, typically overestimated, value of $Q^2$. The
error in the boost leads to a miscalculation of the direction of the Breit 
frame axis. In most cases, this leads to an apparently empty, or at least 
depleted, region of phase space where the scattered quark fragments are 
expected. With improved statistics this has become a significant
effect. To reduce the size of the necessary QED corrections, the new analysis 
utilises more severe cuts on the mass of the total hadronic system, 
$W$, of $W^2>4400\,$~GeV$^2$ both using the scattered lepton variables, 
and separately from the hadronic system variables with the Jaquet Blondel 
method~\cite{JB}.
   
        Following all of the selection procedures there are $20\,810$ events in
the low $Q^2$ sample and $1\,250$ events at high $Q^2$.

        In addition to these event selections there are also cuts
made to reject badly measured tracks. Any tracks in the central chambers 
with transverse momentum below 150 MeV and those in the forward track detectors
 with momentum below 500 MeV are removed. We also remove tracks failing minimum 
requirements on the number of hits in a given chamber and the quality of the
track fit.  In order to have only the 
primary multiplicity, cuts are also made to exclude tracks not originating from
 the interaction vertex. There remains a small excess contribution due to the 
decays of short-lived strange particles. 

        To correct for this excess and for acceptance losses, we utilise the 
DJANGO6~\cite{Dj} Monte Carlo event generator. This combines a LO perturbative 
QCD matrix element calculation and the colour dipole model of hadronisation 
with a calculation of QED radiative effects. Radiative effects remaining after
our event selections have been corrected by comparing the results of Monte
Carlo calculations with and without the inclusion of QED radiative effects.
These corrections are at the $\sim$10\% 
level. The detector response is simulated using a program based
on the GEANT~\cite{Geant} package and the simulated events are reconstructed
and selected using exactly the same analysis chain as with real data.
The total bin-by-bin corrections made throughout this analysis are generally 
well within $\pm$20\% and vary smoothly in any given distribution. 

        The largest individual sources of systematic error for this analysis
originate in possible calibration errors of $\pm$1\% for the BEMC and $\pm$3\% 
for the LAr electromagnetic calorimeter energy scales. These directly affect
the accuracy of Lorentz boosts and give rise to an uncertainty in the number
of tracks of $\sim$5\% at low $Q^2$ and $\sim$8\% at high $Q^2$, irrespective
of other kinematic selection. The corresponding systematic error in the Monte 
Carlo derived 
acceptance correction functions has been estimated to be $\sim$~2\%  using
several different generators~\cite{Le,He}. Visual scans of real and simulated 
data have ascertained that the efficiency of the track detectors is modelled 
to an accuracy of better than 2\% by our Monte Carlo simulations. 
 
        The full experimental details of how the corrected distributions of
this analysis are obtained from the data may be found in reference~\cite{PD}.  

\section{The Breit Frame of Reference}
                                          
The \ep Breit frame is aligned with the hadronic centre of mass (HCM) but 
boosted along a common $z$ direction such that the incident virtual photon has 
zero energy, zero transverse momentum and a $z$ component of momentum $-Q$.  
As with the laboratory frame of reference, we choose the positive $z$ axis to 
be in the direction of the incoming proton.  The negative 
$z$ direction is referred to as the `current' hemisphere of the 
interaction. Our earlier Breit frame analysis~\cite{93BF} showed that
multiplicities in the current region of the Breit frame depend on $Q$ and
not on Bjorken $x$ ($\sim Q^2/W^2$), as opposed to the HCM where multiplicities
depend on the natural scale, $W$~\cite{MD}.  In the na\"\i ve quark parton 
model (QPM) the massless incoming quark has energy $Q/2$ and $z$ component
of momentum $+Q/2$, carrying an approximate fraction $x$ of the proton's 
momentum. After scattering it still has energy $Q/2$, with momentum $-Q/2$. 
By comparison, we thus take the equivalent to the \e+e- centre of mass 
energy, $E^*$, to be $Q$.

        As compared with the HCM the Breit frame current hemisphere is
dominated by the fragments of the struck quark alone; the `spectator' proton 
remnants go entirely into the `target' hemisphere, with much higher momentum. 
There is excellent acceptance for the current hemisphere in the central H1 
detector. For example, even in the highest $Q^2$ interval of this analysis
there is only an $\sim$8\% contribution of tracks from the Forward Track
Detector. The HCM current hemisphere, in contrast, has an energy scale of 
$W/2$ rather than $Q/2$ and generally 
can not be seen in its entirety. This makes comparisons with a 
complete \e+e- interaction hemisphere somewhat easier in the Breit frame.

        When QCD corrections to the parton model are considered, the incident 
parton carries, in general, a proportion of the incident proton's momentum
that is larger than $x$ and has an energy larger than $Q/2$.
Furthermore, since the four-momentum of the photon is fixed the energy seen in 
the current hemisphere may be either greater than or less than $Q/2$. In an 
\e+e- interaction it is also the case that the energy in any given hemisphere 
may be greater than or less than $E^*/2$. The effects of final state radiation 
in \ep DIS and \e+e- interactions are similar, but other LO QCD processes which
affect DIS, initial state QCD radiation and boson gluon fusion (BGF), have no
equivalent in \e+e- annihilation. Note that, to order $\alpha_s$, i.e. 
two exiting partons as well as the spectator system, DIS events may have 
{\it no} energy in the current hemisphere when the incident parton has energy 
above $Q$, other than that due to hadronisation~\cite{Streng}.  Momentum
conservation ensures that this situation never exists in any given hemisphere 
of an \e+e- interaction.

\section{Evolution of the Fragmentation Function}     

        The \ep Breit frame equivalent of the \e+e- scaled hadron momentum   
$x_p = 2p_{hadron}/E^*$ is $x_p = 2p_{hadron}/Q$, where only hadrons in the
current hemisphere are considered.  The event normalised distribution 
$D(x_p,Q^2) = (1/N_{evts})\times dn_{tracks}^{\pm}/dx_p$, the fragmentation 
function, characterises the processes by which partons shower and then 
hadronise. In this paper, the intention is to present the spectra of  
charged particles originating from the primary vertex following the 
fragmentation of light quarks. 

\begin{figure}[!hbt]%otherwise use [hbt], but this removes all restrictions
 \begin{center}
\epsfig{file=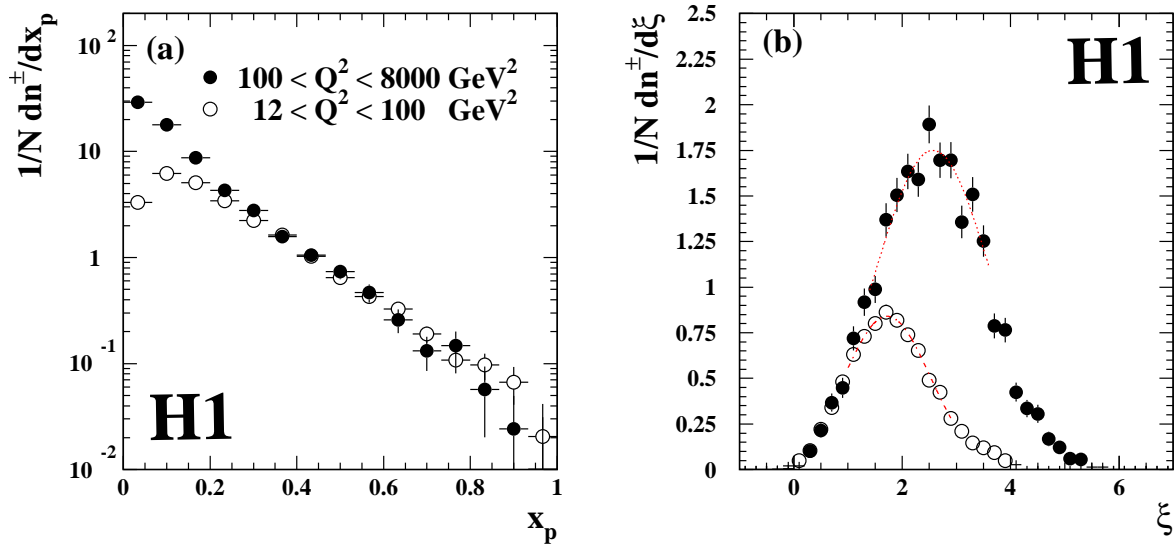,
height=8.5cm}% width setting upsets [h]
 \end{center}
 \caption{\em The current hemisphere fragmentation function as a function
of (a) $x_p$ and (b) $\xi$ shown separately for the low (open
circles) and high (closed circles) $Q^2$ data samples. Statistical and 
systematic errors are added in quadrature. The dashed and dotted lines show 
the results of Gaussian fits.}
\label{Dxxi}
\end{figure}

\begin{figure}[!hbt]%otherwise use [hbt], but this removes all restrictions
 \begin{center}
\epsfig{file=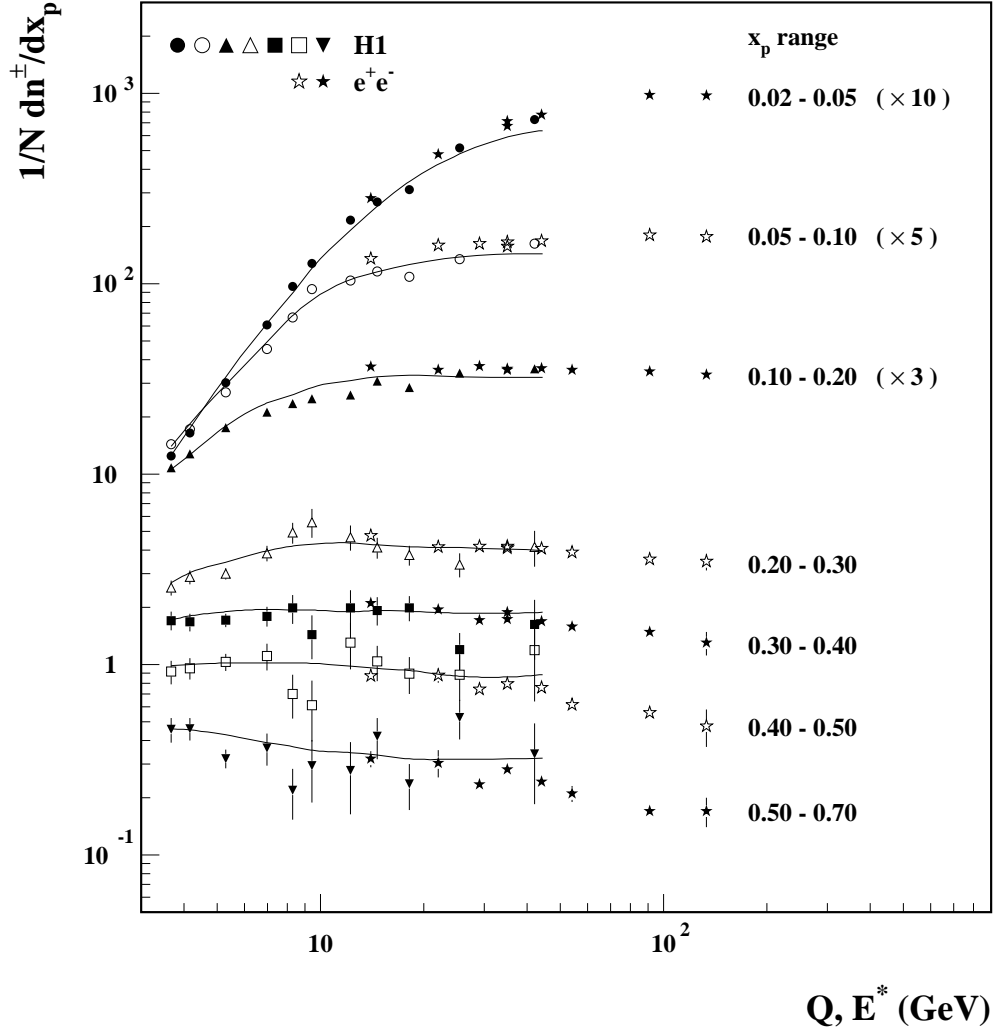,
bbllx=10,bblly=20,bburx=510,bbury=547,
height=15.0cm}% width setting upsets [h]
 \end{center}
 \caption{\em Scaling violations of the fragmentation function as a
function of $E^{*}$ for \e+e- results (starred symbols), and as a function of 
$Q$ for H1 results (all other symbols) each within the same indicated 
intervals of the scaled momentum, $x_p$. Note that the data for the three 
lowest intervals of $x_p$ are multiplied by factors of 10, 5 and 3 respectively
for clarity. Statistical and systematic errors are
added in quadrature. The solid lines show the prediction of a DJANGO Monte 
Carlo calculation.}
\label{Svio}
\end{figure}

        At high enough energies and for light enough final state particles, 
fragmentation functions approximately scale and are ``soft'', rising rapidly 
at small $x_p$ and peaking near $x_p=0$. As the energy of the 
initial parton increases, $D(x_p,Q^2)$ evolves into an even softer function 
with increased probability for low $x_p$ hadrons at the expense of high $x_p$. 
This scaling violation can be seen with the low and high $Q^2$ data 
as a function of $x_p$ in Fig.~\ref{Dxxi}(a) or in Fig.~\ref{Svio} where the 
fragmentation function is plotted as a function of $Q$ for 
different intervals of $x_p$. Typically, in this analysis, the intervals 
of $x_p$ are much greater than the resolution ($\sim$6\% rms). In these figures
the H1 data are compared with equivalent data from \e+e-
experiments~\cite{SVee} plotted as a function of $E^{*}$.  Most published 
\e+e- results refer to full event multiplicities. Here, as elsewhere in this 
analysis, track multiplicity data from \e+e- experiments have been halved to 
correspond to the fragmentation of one timelike quark. Note that H1 as a 
single experiment is able to measure the violation for a wide range of $Q$.
Fig.~\ref{Svio} also displays a prediction from a leading order QCD plus 
hadronisation Monte Carlo (DJANGO6 \cite{Dj}) calculation. The scaling 
violation effect has a similar origin to the scaling violations in structure 
functions, and with next to leading order calculations may be used as a test 
of perturbative QCD~\cite{SVNLO}. 
 
%============================= Figure 1 ==============================

%============================= Figure 2 ==============================

When the fragmentation function is plotted as a function of the variable 
$\xi = \ln (1/x_p)$ the turn-over region is expanded. According to the
expectations of the Modified Leading Logarithmic Approximation (MLLA) and
Local Parton Hadron Duality (LPHD), $D(\xi )$ is Gaussian in the 
neighbourhood of the peak~\cite{Dok}. Even the high statistics of this, 
compared with our earlier, analysis see no significant deviation from this 
Gaussian behaviour, as is clear in Fig.~\ref{Dxxi}(b).   Most track measurement
problems occur with particles of very high or very low momenta and these are 
correlated with low and high $\xi$ respectively. To avoid any dependence on 
these areas the fits of this analysis are restricted to a region within 
one unit of the dimensionless $\xi $ on either side of the mean, but tests in 
which this range is varied by $\pm$20\% show no systematic difference.

\begin{figure}[!hbt]%otherwise use [hbt], but this removes all restrictions
 \begin{center}
\epsfig{file=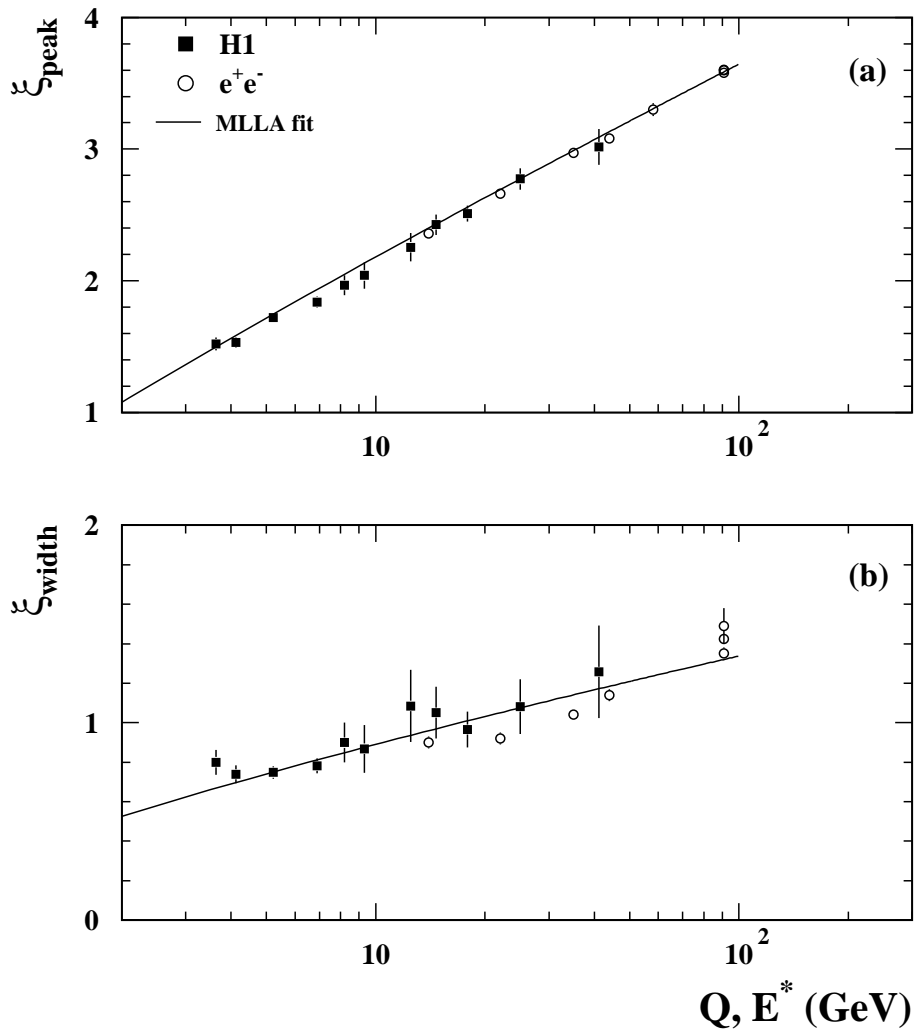,
%QMPHVC$DK300:[Workspace0.Thompson.h1dat]mlla.eps,
bbllx=10,bblly=30,bburx=530,bbury=570,height=14.3cm}% width setting upsets [h]
 \end{center}
 \caption{\em 
        H1 results (solid symbols, with statistical and systematic errors
combined in quadrature) showing the evolution of (a) the peak and 
(b) the width of the fragmentation function as a function of $Q$ compared with 
\e+e- results (open symbols) as a function of the
centre of mass energy, $E^*$. The solid line is a fit to MLLA/LPHD expectations.
}
\label{PWQ}
\end{figure}

                The evolution of the fragmentation function may be summarised 
by the $Q$, or $E^*$, dependence of the peak and width (dispersion)
values of the fitted Gaussian as is shown in Fig.~\ref{PWQ} and in 
Table~\ref{TPWQ}. It is notable that the systematic effects discussed in 
section 3 contribute only a total 2\% uncertainty to the peak value and have 
no significant effect on the width measurement. The results are
compatible with those published earlier, which had considerably lower precision,
as well as with those of an analysis~\cite{ZEUS} by the ZEUS collaboration.
The results are also compatible with various \e+e- 
experiments~\cite{SVee,OPAL,ALEPH}, where the relevant evolution variable is 
the centre of mass energy, $E^*$. For this comparison and for the consideration
of multiplicities in the next section, these published \e+e- data have 
been re-fitted by us to be in a directly comparable form.  As well as the 
previously mentioned factor of two to convert \e+e- event multiplicities to 
quark multiplicities, we also reduce published total multiplicity results by 
8\% to account for $K^0$ and $\Lambda$ decay tracks~\cite{K0,Acton}, and by a 
variable factor below 3\% to correct for the 
increased multiplicity from $b$ quark fragmentation~\cite{Bmult} in {\e+e-}.

%============================= Figure 3 ==============================
%============================= Table 1 ==============================
\begin{table*}
\begin{center}
\begin{tabular}{|c||c|c|c||c|c|c|} \hline 
\multicolumn{1}{|c||}{\bf {\mQ}} &
\multicolumn{3}{|c||}{\bf {Total Current Hemisphere}} & 
\multicolumn{3}{|c|}{\bf {Energy Flow Selected}} \\ 
%\hline \hline 
\cline{2-7}
\cline{2-7}
\multicolumn{1}{|c||}{ (GeV)} 
&\multicolumn{1}{|c|}{\avn}
&\multicolumn{1}{|c|}{Peak} 
&\multicolumn{1}{|c||}{Width}
&\multicolumn{1}{|c|}{\avn}
&\multicolumn{1}{|c|}{Peak} 
&\multicolumn{1}{|c|}{Width} \\ \hline \hline  
3.63 &  % {12 $\to$ 15} & 
{1.18$\pm$0.07}&{1.52$\pm$0.05}&{0.80$\pm$0.06}&   
{1.77$\pm$0.11}&{1.51$\pm$0.05}&{0.76$\pm$0.07}\\ \hline \hline 
4.12 &  %{15 $\to$ 20} &
{1.38$\pm$0.08}&{1.53$\pm$0.04}&{0.74$\pm$0.05}& 
{1.85$\pm$0.11}&{1.51$\pm$0.03}&{0.68$\pm$0.04} \\ \hline \hline
5.23 &  %{20 $\to$ 40} &
{1.62$\pm$0.08}&{1.72$\pm$0.03}&{0.75$\pm$0.03}&
{2.13$\pm$0.11}&{1.70$\pm$0.03}&{0.76$\pm$0.03} \\ \hline \hline
6.90 &  %{40 $\to$ 60} &
{2.19$\pm$0.12}&{1.84$\pm$0.04}&{0.78$\pm$0.04}&
{2.60$\pm$0.15}&{1.83$\pm$0.04}&{0.77$\pm$0.05} \\ \hline \hline
8.20 &  %{60 $\to$ 80} &
{2.59$\pm$0.16}&{1.97$\pm$0.08}&{0.90$\pm$0.10}& 
{3.10$\pm$0.20}&{1.96$\pm$0.07}&{0.87$\pm$0.10} \\ \hline \hline
9.32 &  %{80 $\to$ 100} &
{3.16$\pm$0.25}&{2.04$\pm$0.10}&{0.87$\pm$0.12}& 
{3.49$\pm$0.29}&{2.06$\pm$0.09}&{0.84$\pm$0.11} \\ \hline \hline
12.5 &  %{100 $\to$ 175} &
{3.65$\pm$0.32}&{2.26$\pm$0.11}&{1.09$\pm$0.18}& 
{3.93$\pm$0.35}&{2.23$\pm$0.10}&{1.03$\pm$0.17} \\ \hline \hline
14.6 &  %{175 $\to$ 250} &
{4.13$\pm$0.34}&{2.43$\pm$0.08}&{1.05$\pm$0.13}&
{4.45$\pm$0.30}&{2.41$\pm$0.07}&{1.04$\pm$0.13} \\ \hline \hline
17.9 &  %{250 $\to$ 450} &
{4.05$\pm$0.34}&{2.51$\pm$0.06}&{0.97$\pm$0.09}& 
{4.28$\pm$0.36}&{2.49$\pm$0.06}&{0.98$\pm$0.10} \\ \hline \hline
25.0 &  %{450 $\to$ 1000} &
{5.23$\pm$0.43}&{2.77$\pm$0.08}&{1.08$\pm$0.14}&
{5.28$\pm$0.44}&{2.80$\pm$0.10}&{1.18$\pm$0.18} \\ \hline \hline
41.2 &  %{1000 $\to$ 8000} &
{7.32$\pm$0.65}&{3.02$\pm$0.14}&{1.26$\pm$0.23}&
{7.19$\pm$0.65}&{2.95$\pm$0.14}&{1.19$\pm$0.20} \\ \hline
\end{tabular}
\end{center}
\caption{ \sl The average charged multiplicity, and the peak ($\xi_{peak}$) 
and width ($\xi_{width}$) of the fragmentation function as a function of $Q^2$,
for the total current hemisphere of the Breit frame both with and without the 
energy
flow selection discussed in section 6. The errors are the sum of statistical 
and systematic uncertainties in quadrature.}
\label{TPWQ}
\end{table*}                   

        Assuming gluon coherence MLLA/LPHD predicts a dependence of the 
evolution of the peak and width of the distribution on the dimensionless
 variable $Y = \ln (Q/\Leff) $, where $\Leff $ is a scale parameter determining
a presumed cut--off of a parton shower. The prediction~\cite{Dok} gives the 
peak, 
$\xi_{peak}$ and the width, $\xi_{width}$, of the Gaussian approximation to be 
$$\xi_{peak}= 0.5Y + c_2\sqrt{Y} + {\cal K}$$
$$\xi_{width}= \sqrt{Y^{3\over 2}/2c_1}{\rm ,}$$
        where $c_1=\sqrt{36N_c/b}$ and $c_2=B\sqrt{{{1}\over{16}}b/N_c}$ with
$b={{11}\over{3}}N_c-{{2}\over{3}}N_f$ and 
$B=({{11}\over{3}}N_c+{{2}\over{3}}N_f/N_c^2)/b$ are constants dependent only 
on the number of contributing colours, $N_c$, and flavours, $N_f$ within the
parton shower. For ease of comparison, we follow \cite{OPAL,ALEPH} in the 
assumption that this will be dominated by the light quarks and set $N_f=3$. 
The term ${\cal K}$ contains higher order corrections and is expected to be 
roughly constant and of order 1. The result of a simultaneous fit 
($\chi^2$/NDF of 10/20) to the peak and width values obtained from the present 
H1 data alone is shown as a solid line in Fig.~\ref{PWQ} and yields the
values $\Leff=0.21\pm 0.02$~GeV and ${\cal K}=-0.43\pm 0.06$, in agreement with 
the values $\Leff=0.21\pm 0.02$~GeV and ${\cal K}=-0.32\pm 0.06$ obtained in an
analysis~\cite{OPAL} of the $\xi_{peak}$ evolution of combined \e+e- data.

\section{The Average Charged Particle Multiplicity}     

\begin{figure}[!hbt]%otherwise use [hbt], but this removes all restrictions
 \begin{center}
\epsfig{file=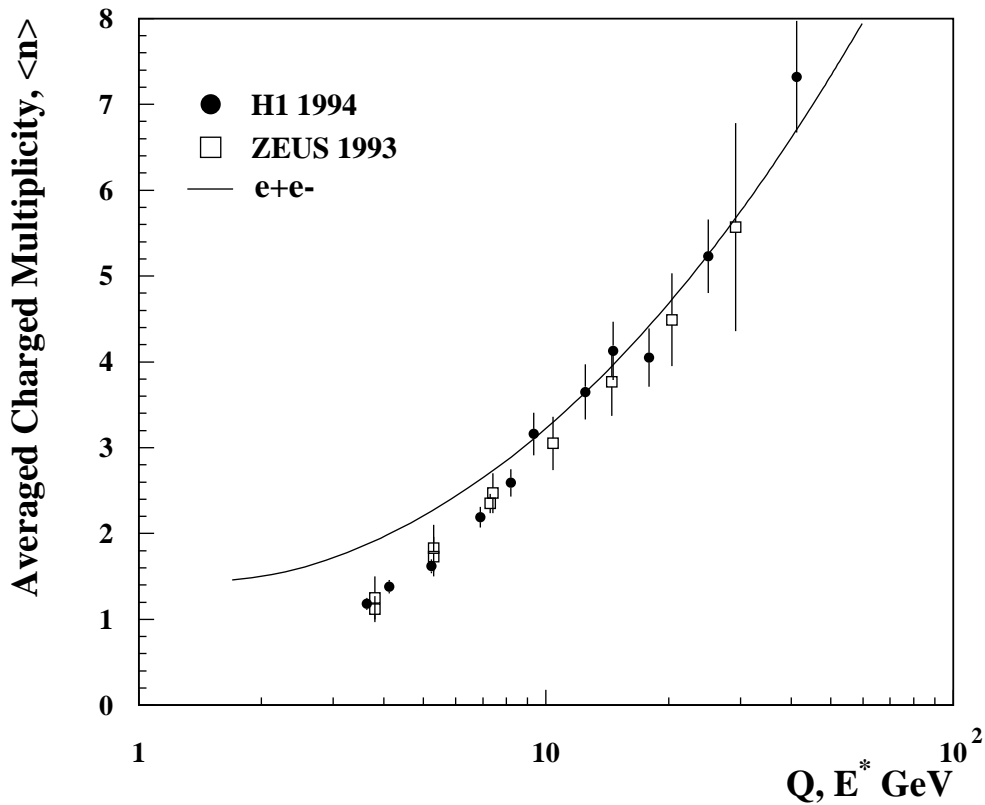,
bbllx=50,bblly=30,bburx=550,bbury=470,height=10.5cm}% width setting upsets [h]
% 17cm is too big for this pix.
 \end{center}
 \caption{\em
        Average charged multiplicity in the current hemisphere of the
Breit frame as a function of $Q$ for this analysis (solid circles) and for 
the ZEUS 1993 analysis (open squares). Statistical errors and systematic errors
are added in quadrature.
The curve is a fit to many \e+e- results as a function of the centre of mass 
energy, $E^*$.
}
\label{Avn}
\end{figure}

        The area underneath the fragmentation function is the average
charged multiplicity and this is given, after all corrections, in 
Table~\ref{TPWQ}. It is shown as a function of $Q$ in Fig.~\ref{Avn}, with ZEUS 
results~\cite{ZEUS} and with a curve which accurately 
parameterises~\cite{Acton} the 
average charged multiplicity from many \e+e- experiments as a function of $E^*$.
The error associated with this fit at any given energy is at the percent level.
%============================= Figure 4 ==============================

        There are predictions~\cite{Bass} that both the absolute  
average charged DIS multiplicity and the shape of the evolution with $Q^2$ 
should agree with that of \e+e- experiments. As previous results indicated, 
although there
was very good agreement between H1 and ZEUS~\cite{ZEUS} analyses and those of 
\e+e- at high $Q (E^*)$, the DIS analyses give smaller average charged 
multiplicities at low values of $Q (E^*)$. The new data confirm that this is a 
large effect. 

        The discussion of section 4 indicated that the LO processes
which are present in \ep but not in \e+e- interactions can produce a
depletion of the current region of the Breit frame. Pure
QPM interactions {\em or} events with only final state radiation (i.e. \e+e- 
like events) might both be expected to have a current region hadronic state 
characteristic of a quark recoiling with energy $\sim Q/2$ and with only
a $z$ component of momentum. BGF events with a depleted or empty current 
hemisphere 
or events with significant initial state QCD radiation would, on the contrary,
have a current region hadronic final state with significantly less energy
at an angle to the $z$ axis. However, hadronisation effects might well
mask these na\"\i{ve} expectations.

\begin{figure}[!hbt]%otherwise use [hbt], but this removes all restrictions
 \begin{center}
\epsfig{file=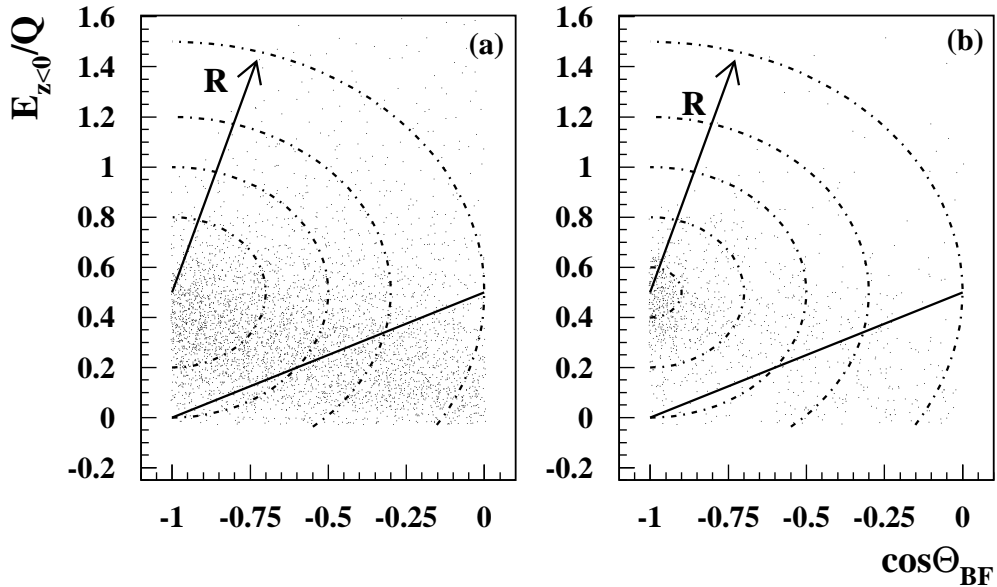,
bbllx=100,bblly=40,bburx=550,bbury=370,height=8.0cm}% width setting upsets [h]
 \end{center}
 \caption{\em
        The total energy of the summed calorimeter cluster four--momentum
vectors in the current hemisphere of the Breit frame is plotted as a fraction
of the event $Q$ against the polar angle of the resultant vector, for (a) the
low $Q^2$ and (b) the high $Q^2$ data sample. The annuli define the $R$ 
variable for Fig.~\ref{nvsR} and the solid line indicates the 
Breit frame energy selection referred to in the text.
}
\label{Etheta}
\end{figure}

        In order to investigate this experimentally, we add the four--momenta of
all calorimeter energy clusters in the current region of the Breit frame. The 
energy component of the resultant four--momentum vector is defined to be 
$E_{z<0}$ and the angle it makes with the $z$ axis to be $\Theta_{BF}$. We 
then plot 
$E_{z<0}/Q$ against $\cos \Theta_{BF}$, as shown in Fig.~\ref{Etheta}(a,b) 
for the low  and high $Q^2$ samples. From the above discussion we would expect 
that \e+e- like events might cluster near ($\cos \Theta_{BF}=-1, 
E_{z<0}/Q=1/2$) in this plot, whereas events from the other LO processes would 
either not appear in this plot or be spread away from this point, typically
towards lower energies and larger angles in the current hemisphere, that is to 
the lower right-hand corner of this plot.
 The high $Q^2$ sample, where $\alpha_s$ is small, $x$ is large, and the proton
has a 
much smaller gluon content, shows obvious clustering near the (-1,$0.5$) point,
while this is much less evident at low $Q^2$. 

\begin{figure}[!hbt]%otherwise use [hbt], but this removes all restrictions
 \begin{center}
\epsfig{file=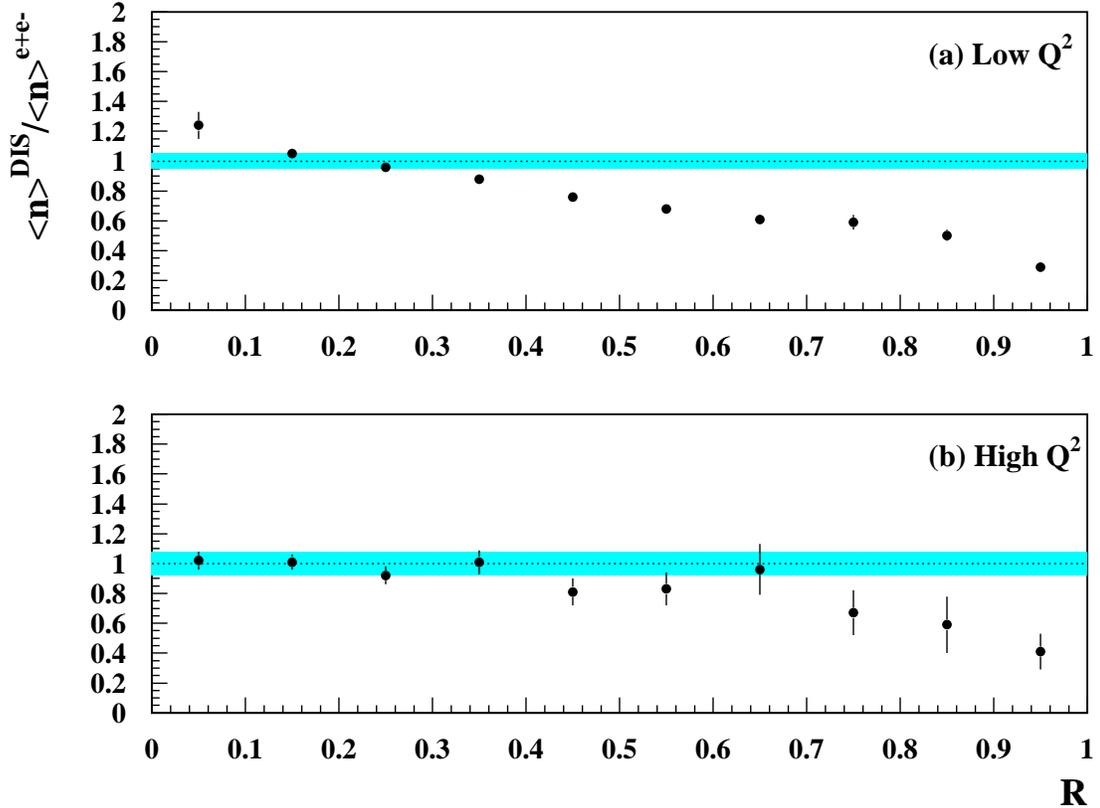,
bbllx=70,bblly=50,bburx=550,bbury=460,height=10.0cm}% width setting upsets [h]
 \end{center}
 \caption{\em   The H1 average charged multiplicity as a function of the radius 
defined in the text normalised to that of the single hemisphere expectation for
\e+e- events  for (a) the low $Q^2$ and (b) the 
high $Q^2$ data sample. The H1 points have statistical error bars only. 
The shaded regions show the total systematic error.
}
\label{nvsR}
\end{figure}

        Monte Carlo simulations qualitatively confirm the expectations but it
is hard to be quantitative, both because of the difficulty in strict 
definitions of LO QCD event classes and because experimental resolution has a 
large contribution to event migration on this plot. Nevertheless,
 it is possible~\cite{DK} to investigate the effect of choosing ever 
more \e+e- like events by selecting according to the variable $R$, where 
$R^2 = (E_{z<0}/Q-0.5)^2 + (\cos \Theta_{BF}+1)^2$.  In order to compare with
\e+e- results, we divide $\left< n(R) \right>$, the average multiplicity of
events in an annulus of width $\Delta R =0.1$ at radius $R$, by the equivalent 
single--hemisphere \e+e- multiplicity. There is a very small variation of \mQ 
with $R$, so this \e+e- multiplicity is evaluated at an energy 
corresponding to the average value of $Q$ of the H1 data in that annulus. The 
behaviour of this ratio as a function of $R$ can be seen, with statistical 
error bars only, in Fig.~\ref{nvsR}(a,b). The systematic errors corresponding
to the shaded region are as discussed in section 3. Events with a totally 
empty current hemisphere have zero multiplicity but no defined value of $R$, 
and are thus not included in these plots. As $R\to 0$ the average charged 
hadronic multiplicity {\it increases}, possibly even beyond the corresponding 
\e+e- values, although it is not inconsistent with \e+e- values given the 
systematic errors involved.  Our Monte Carlo simulations also show this general 
behaviour.
%============================= Figure 5 ==============================
                                                                        
%============================= Figure 6 ==============================

        The above result and discussions motivate a cut to remove extreme low 
energy and off--axis events. This is shown as a solid line in 
Fig.~\ref{Etheta}(a,b) joining the points (-1,0) and (0,0.5) in this plane and 
is termed the `Breit frame energy flow selection' in the rest of this paper. It 
produces samples of $11\,735$ events and $1\,005$ events at low and high $Q^2$ 
respectively, which are less affected by the low--order QCD processes peculiar
to DIS.
This selection is by no means as exact as the technique of letting
$R\to 0$ but in circumstances where the data are already binned in other
variables, e.g. energy spectra or multiplicity distributions, gives an 
indication of whether or not distributions move in the direction of closer 
similarity between this data and \e+e- data.

        As is clear from Table~\ref{TPWQ}, the peak and width values of the 
fragmentation function are insensitive to this energy flow selection procedure.
 The average multiplicity, however, is sensitive and as expected increases
with this selection, especially at low $Q^2$.

\section{Invariant Energy Spectra}

The evolution of the fragmentation function at low energies, or high values of 
$\xi$, may be better studied using the event--normalised ``invariant'' 
spectrum, $(1/N_{evts})\times Edn_{tracks}^{\pm}/d^3p$. According to the 
predictions of perturbative QCD based on the MLLA and LPHD, the 
hadronic spectrum at low momentum should be nearly independent of the energy of 
the parent parton~\cite{Lup}. Here we follow \cite{Lup} in taking 
$d^3p=4\pi p^2_{hadron}dp_{hadron}$ 
and by computing the energy, $E$, of each track as if it had a mass 
$Q_0(=$270~MeV$\approx \Lambda_{QCD})$ given by a presumed cut-off mass for the 
parton shower, i.e. $E = \sqrt{Q_0^2 + p_{hadron}^2}$. The data are corrected 
for acceptance with the same bin-by-bin method described in section 3. The
correction function is both smooth and close to unity, having an average value
of around 1.15. Although there is always a loss of track acceptance at low 
laboratory momentum the boost reduces this loss in the Breit frame. This means
that \ep experiments can get closer to the critical low energy limit without
serious loss of acceptance, than can \e+e- experiments.
                
\begin{figure}[p]%otherwise use [hbt], but this removes all restrictions
 \begin{center}
\epsfig{file=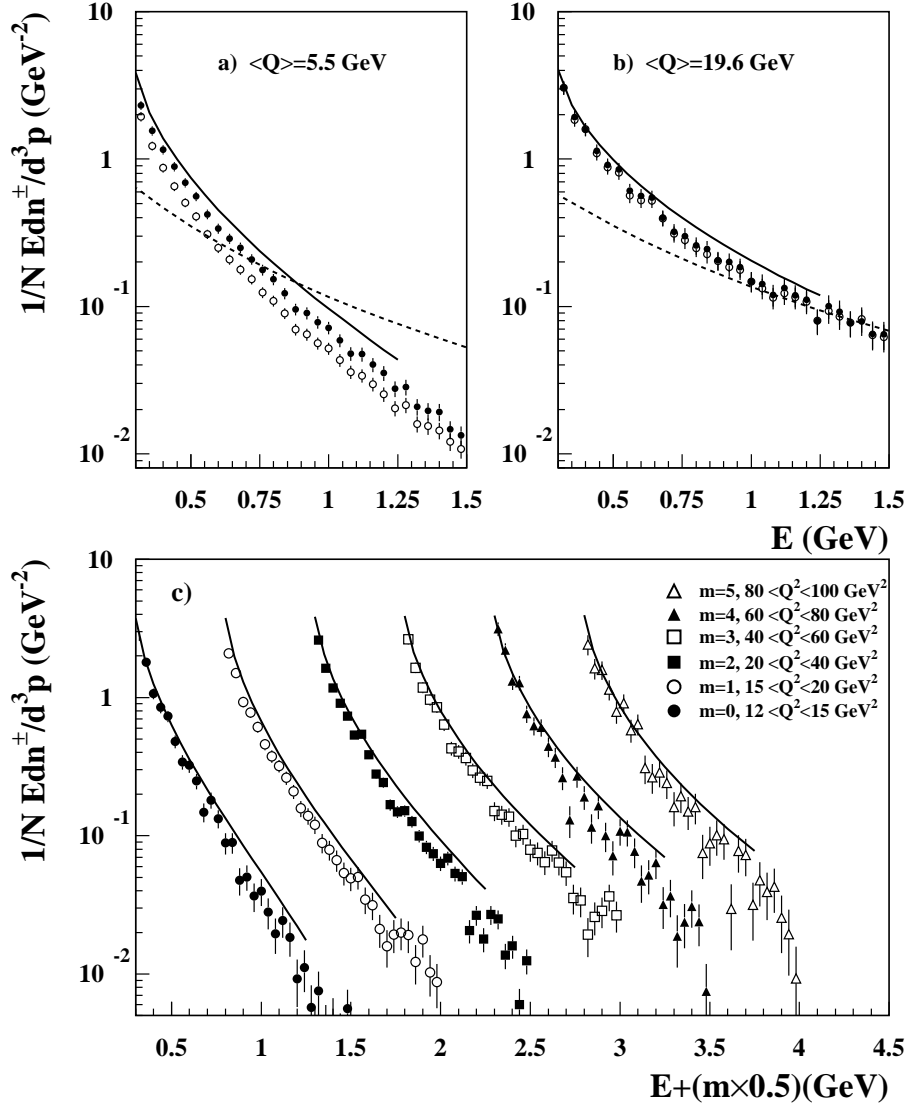,
bbllx=0,bblly=0,bburx=638,bbury=794,height=17.0cm}% width setting upsets [h]
 \end{center}
 \caption{\em The invariant charged hadron energy spectrum in the current 
hemisphere (a) at low $Q^2$ and (b) at high $Q^2$. The data for all events are 
shown as open circles and those utilising the Breit frame energy flow selection 
as solid circles.
The solid line is the prediction of MLLA/LPHD and the dashed line is the 
corresponding expectation for a non-running coupling constant, $\alpha_s$.
In (c) the low $Q^2$ data are subjected to the energy flow selection but 
subdivided into the six indicated intervals of $Q^2$ and plotted with a 
regular incremental spacing of 0.5~GeV on the abscissa. The solid line
MLLA/LPHD expectation is calculated at the \mQ for each distribution.
The error bars show the sum of statistical and systematic
errors added in quadrature.
}
\label{Inv}
\end{figure}

The results are shown in Fig.~\ref{Inv}(a,b) where the indicated errors contain 
contributions from statistical and (negligible) momentum measurement effects 
as well as systematic effects that have already been discussed and which grow
slightly with track energy. The solid lines show MLLA/LPHD predictions at 
the relevant $Q$ values for a simple quark behaviour using the calculations 
of~\cite{Lup} which the authors show are in agreement with \e+e- data. It is 
clear that at low $Q^2$ these predictions for the behaviour of quark fragments
are in disagreement with the data but are significantly closer to data 
subject to the Breit frame energy flow selection described in the previous 
section. This selection makes little difference at high $Q^2$, where agreement
is better anyway. The QCD predictions are very sensitive to the `running' of the
strong coupling constant, $\alpha_s$, as the energy scale alters in the parton
shower. A prediction is shown for the same average energy, normalised to 
interpolated $\left<n^{\pm}\right>$ multiplicities in \e+e- experiments, but 
utilising a fixed value of $\alpha_s$ (set to 0.215, a value which roughly 
describes the energy dependence of $\left<n^{\pm}\right>$ and the slope of the 
first moment of the energy spectrum in \e+e- experiments~\cite{Lup2}).  The 
data show that such an assumption is clearly untenable. This calculation is
only in leading order so this cannot yet be turned 
around to compute a meaningful value of $\Lambda_{QCD}$.
Data subject to the energy flow selection are re-plotted in six intervals of
$Q^2$ in Fig.~\ref{Inv}(c) in order to display the clear evidence for a common 
limit at low particle energy independent of event momentum transfer. 

%============================= Figure 7 ==============================

\section{Charged Particle Multiplicity Distributions}

        Charged hadron multiplicity  distributions are considerably more 
difficult to measure than average values. In this analysis we have tested a
number of ways of extracting %~\cite{Mnote} 
true distributions from observed distributions. The best method is
found to be a simple bin-by-bin correction procedure on the multiplicity 
distribution. This has, however, the problem of not modelling correlations
which then gives the method a dependence on the Monte Carlo generator.

        We tried two further methods to solve this difficulty.
It is straightforward to use a Monte Carlo simulation to form a generated,
or `true', to reconstructed distribution transfer matrix, which is
physics-generator independent. Inversion of this matrix in order to extract a 
true 
distribution from a raw input distribution has many problems due to the small 
number of high multiplicity events, resulting in the matrix becoming singular
and unstable. We have also tried a method~\cite{MD,pvm} of
approximating the inverted matrix with its inverse and then re-weighting the 
input Monte Carlo distribution in successive approximation iterations to 
minimise dependence on the initial generated distribution. 

        The systematic errors of all three techniques have been estimated by 
degrading the output distributions back to the `raw' (reconstructed) data level 
using the reliable `generated to reconstructed' matrix. This can then be 
compared with the original, real input distribution.  We find that the larger 
dependence on the generated distribution of the bin-by-bin correction method
is still offset by overall smaller systematic errors and the ability to get 
meaningful corrections at high multiplicity when the statistics are low. We 
thus choose this method to present our results. Quantitatively, the average 
correction made per bin is 16\%(32\%) at low (high) $Q^2$ with an error of 
method assessed as 10\%(17\%). 

\begin{figure}[p]%otherwise use [hbt], but this removes all restrictions
 \begin{center}
\epsfig{file=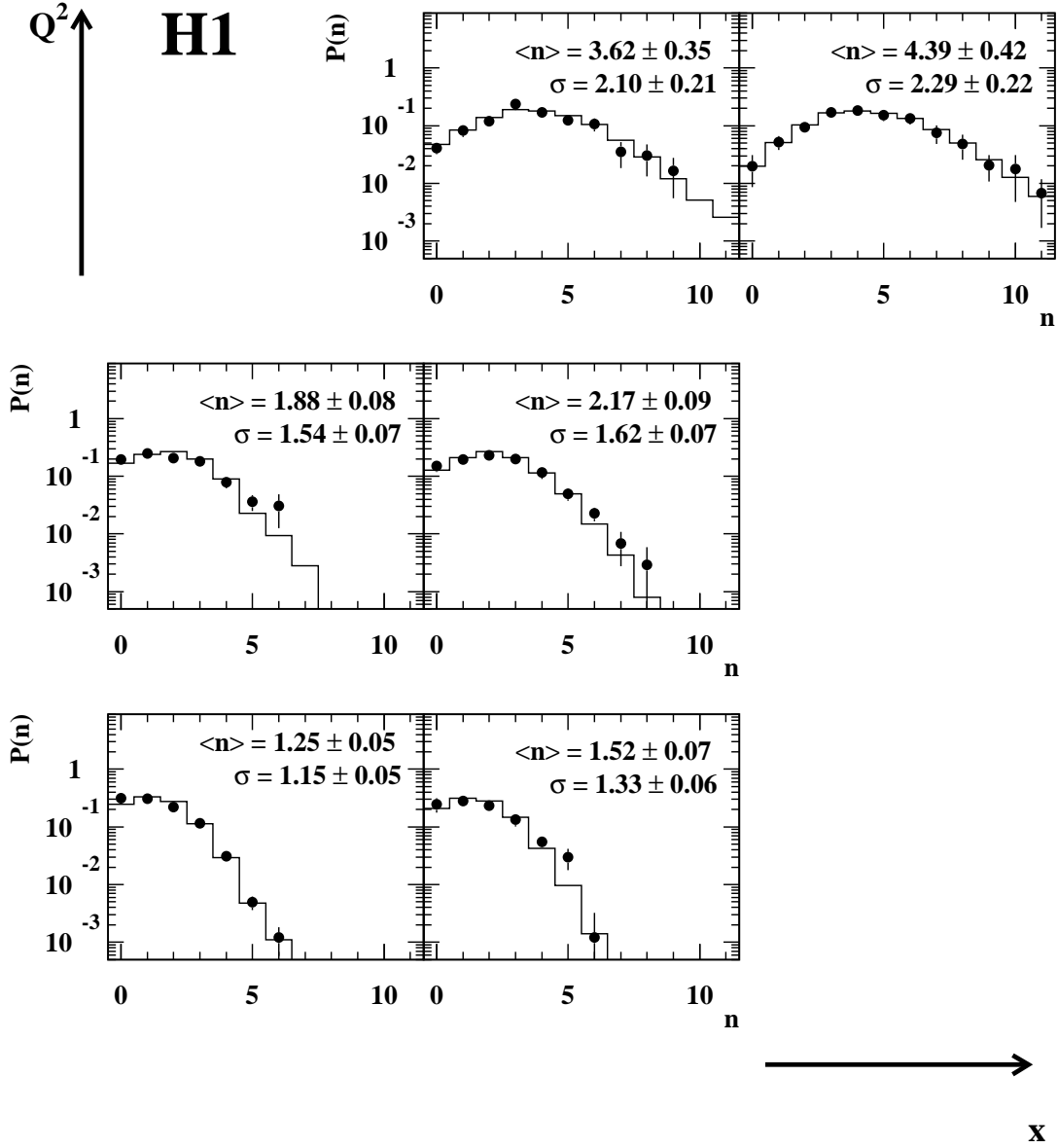,
bbllx=20,bblly=40,bburx=550,bbury=690,height=16.5cm}% width setting upsets [h]
 \end{center}
 \caption{\em The event-normalised charged hadron multiplicity 
distribution in the current region of the Breit frame, with statistical and 
systematic errors added in quadrature.  
Moving from bottom to top the distributions are for $12<Q^2<30$ GeV$^2$,
$30<Q^2<80$ GeV$^2$ and $100<Q^2<500$ GeV$^2$, and from left to right the ranges
in Bjorken-$x$ are $6\cdot 10^{-4}<x<2\cdot 10^{-3}$, 
$2\cdot 10^{-3}<x<1\cdot 10^{-2}$ and $1\cdot 10^{-2}<x<2\cdot 10^{-1}$.
The average value of $Q$ and $x$ for each interval is given in
Table~\ref{nmult}. The average multiplicity and dispersion are shown at the
top of each distribution. The histograms show the prediction
 of a DJANGO Monte Carlo calculation.
}
\label{mult2}
\end{figure}

        The resulting multiplicity distributions are 
given in Table~\ref{nmult} and shown in Fig.~\ref{mult2} in the form of 
$P(n)$, the probability of observing a charged multiplicity $n$.
 These results are for the multiplicity distribution in the current hemisphere 
of {\it all} events and do not include any energy flow selection. The bins of 
$x$ and $Q^2$ are exclusive but, because of the event selections, are not 
totally occupied at the edges. 
This leads to only minor discrepancies in the average values of $x$ and $Q^2$ 
between neighbouring bins. In general the observation is that the results are
compatible with a gradual shift to higher multiplicities with increasing $Q^2$
but with only a small change as a function of $x$ which might be due to minor 
bin edge effects. It is interesting to note the many zero-charged multiplicity 
events, especially at low $Q^2$ where we might expect higher-order QCD
processes to dominate.
%============================= Figure 8 ==============================
%============================= Table 2 ==============================
\begin{table*}
\begin{center}
\begin{tabular}{|c|c||c|c|c|c|} \hline 
{\mQ} & {$<x>$} & 
     {\bf $P(0)$} &{\bf $P(1)$} &{\bf $P(2)$} &{\bf $P(3)$} \\
\cline{3-6}
{(GeV)} & {$(\times 10^3)$}&{\bf $P(4)$} &{\bf $P(5)$} 
& {\bf $P(6)$} &{\bf $P(7)$} \\
\cline{3-6}
        &       & {\bf $P(8)$} &{\bf $P(9)$} & {\bf $P(10)$} &{\bf $P(11)$} \\ 
\hline \hline 
{4.15} & {1.26} 
        & $0.312\pm 0.025$   & $0.309\pm 0.016$ & $0.219\pm 0.027$& $0.116\pm 0.009$  \\
\cline{3-6}
&       &$0.031\pm 0.005$ & $0.005\pm 0.001$    & $0.0012\pm 0.0006$ &  -    \\
\cline{3-6}
&       &            -  &         -         &  -               &  -   \\ \hline
{4.53} & {2.55}
        & $0.246\pm 0.070$   & $0.281\pm 0.025$ & $0.233\pm 0.042$& $0.133\pm 0.031$ \\
\cline{3-6}
&       &$0.055\pm 0.011$ & $0.030\pm 0.012$    & $0.0012\pm 0.0020$ &   -   \\
\cline{3-6}
&       &            -  &         -         &  -               &  -   \\ \hline
{6.20} & {1.58} 
        & $0.194\pm 0.023$   & $0.250\pm 0.026$ & $0.205\pm 0.025$& $0.182\pm 0.023$ \\
\cline{3-6}
&       & $0.079\pm 0.017$ & $0.036\pm 0.011$ & $0.031\pm 0.018$ &       -    \\
\cline{3-6}
&       &            -  &         -         &  -               &  -   \\ \hline
{7.04} & {4.10} 
        & $0.150\pm 0.031$   & $0.195\pm 0.013$ & $0.229\pm 0.025$& $0.199\pm 0.020$ \\
\cline{3-6}
&       & $0.116\pm 0.026$ & $0.049\pm 0.012$   & $0.023\pm 0.006$& $0.007\pm 0.004$ \\
\cline{3-6}
&       & $0.003\pm 0.003$&            - &        -   &  -  \\ \hline
{14.5} & {6.82} 
        & $0.041\pm 0.009$   & $0.083\pm 0.019$ & $0.119\pm 0.024$& $0.236\pm 0.049$ \\
\cline{3-6}
&       & $0.169\pm 0.024$ & $0.124\pm 0.024$ & $0.106\pm 0.027$& $0.035\pm 0.017$ \\
\cline{3-6}
&       & $0.030\pm 0.017$ & $0.017\pm 0.011$   &        -          &  -  \\    \hline
{17.3} & {2.34} 
        & $0.020\pm 0.011$   & $0.052\pm 0.014$ & $0.093\pm 0.018$& $0.171\pm 0.031$ \\
\cline{3-6}
&       & $0.184\pm 0.026$ & $0.151\pm 0.032$ & $0.135\pm 0.029$& $0.075\pm 0.026$ \\
\cline{3-6}
&       & $0.048\pm 0.022$ & $0.021\pm 0.010$ & $0.018\pm 0.013$ & $0.007\pm 0.005$ \\ 
    \hline
\end{tabular}
\end{center}
\caption{ \sl The probability per event, $P(n)$, of observing a charged 
hadronic multiplicity of $n$ in the current hemisphere of the Breit frame 
as a function of $Q$ and $x$. The errors are the sum of statistical 
and systematic errors in quadrature.}
\label{nmult}
\end{table*}                   
%============================= Figure 9 ==============================
\begin{figure}[p]%otherwise use [hbt], but this removes all restrictions
 \begin{center}
\epsfig{file=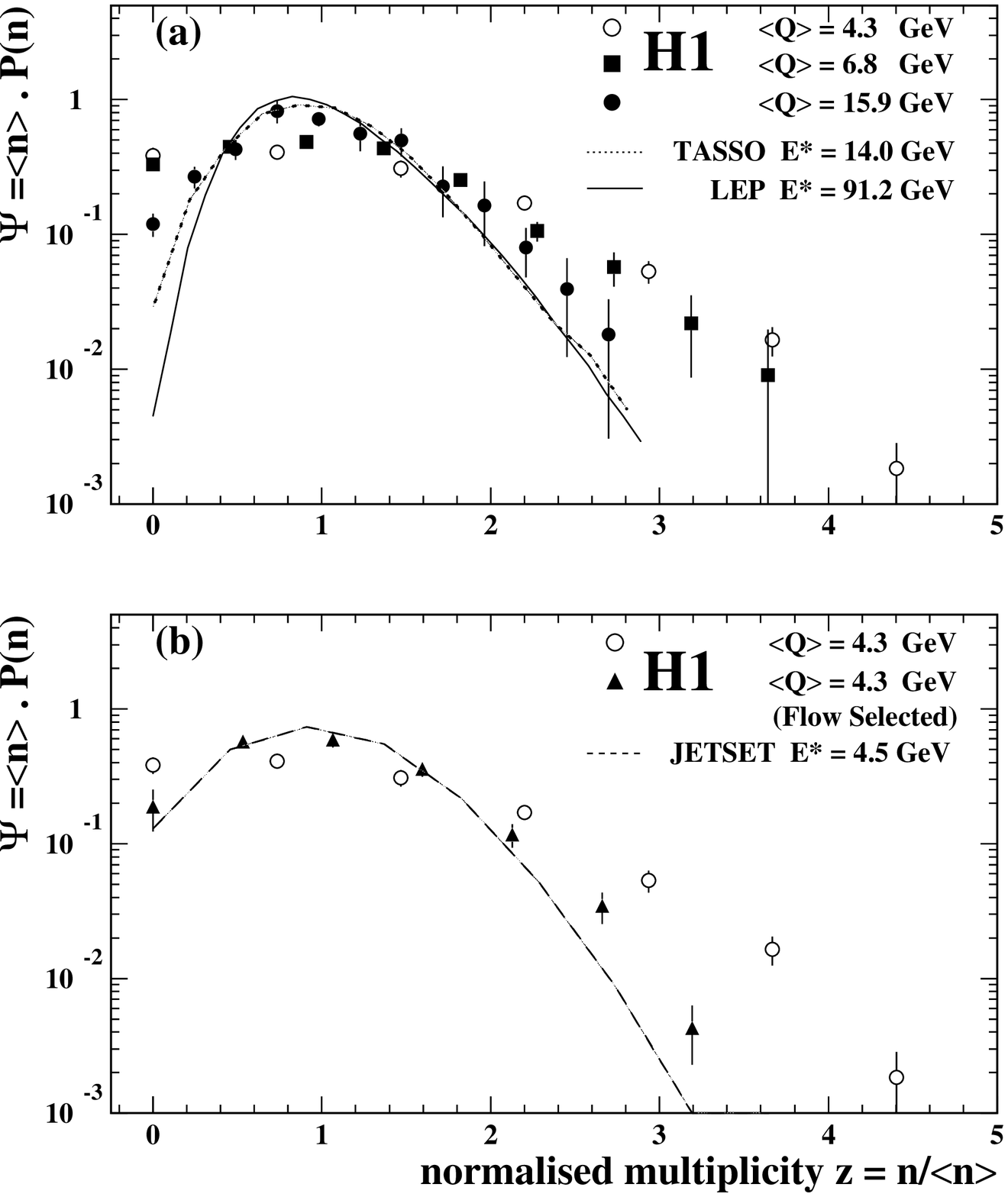,bbllx=20,bblly=20,bburx=500,bbury=670,height=15.5cm}
% width setting upsets [h]
 \end{center}
 \caption{\em 
(a) The charged hadronic multiplicity distribution for H1 data 
expressed in terms of the KNO variables for $12<Q^2<30$~GeV$^2$ (open circles), 
$30<Q^2<80$~GeV$^2$ (squares) and $100<Q^2<500$~GeV$^2$ (solid circles).
(b) A repeat of the $12<Q^2<30$~GeV$^2$ H1 data with (solid triangles) and 
without (open circles) the Breit frame energy flow selection.
Statistical and point to point systematic error bars are added in quadrature 
in both figures. The dotted and solid lines represent the single hemisphere 
KNO functions for \e+e- data at $E^*=$14 and 91 GeV respectively and the
dashed line shows the result from the \e+e- JETSET Monte Carlo at an $E^*$
value close to the $Q$ value of the \ep data. 
}
\label{KNO}
\end{figure}

        Integrating over $x$ gives smaller statistical errors and a 
greater reach in multiplicity. The evolution with $Q$ of the shape of the 
multiplicity distribution may be studied by normalising with the 
average charged hadron multiplicity to give the KNO~\cite{KNO} form 
$\Psi(z)=\mavn \times P(n)$, where $z=n/\mavn$.  Equivalent 
single-hemisphere \e+e- distributions~\cite{Acton,KNOT,KNOD}  exhibit clear 
scaling in this variable but only for high ($E^{*}> 20$~GeV) energies.
In an analysis~\cite{MD} of multiplicity distributions in a limited region
of the hadronic centre of mass frame, the H1 collaboration has also observed 
good agreement with a common distribution as $W$ increases between 80 and 
220~GeV. The results of this
analysis in the Breit frame, shown in Fig.~\ref{KNO}(a), are for $Q$ values
much lower than the $W$ values and show a clear violation of KNO scaling, 
as also reported by the ZEUS collaboration~\cite{ZEUS}. For reference, we
display two sets of lines joining \e+e- data at an energy close to our largest
$Q$ range~\cite{KNOT} and also at LEP energies~\cite{Acton}. As with \e+e-
data, the H1 low $Q^2$ distribution starts with a flatter curve and a large 
contribution from low multiplicities. Then, as $Q^2$ increases, the data get
closer to the high energy scaling \e+e- distributions. 

         In Fig.~\ref{KNO}(b) the distribution for low $Q^2$ events is compared 
with results from the \e+e- JETSET~\cite{Jetset} parton shower model, using 
parameter settings determined by the DELPHI collaboration~\cite{Tune} at an 
energy set to be close to the mean value of the low $Q^2$ data. The 
data
distribution is again much flatter but a better agreement is achieved when
using the Breit frame energy flow selection discussed in section 6.
This KNO scaling violation is also seen in \ep DIS Monte Carlo generated 
events, where it is to a large extent caused by leading 
order perturbative QCD events, especially BGF events, which, as has already 
been noted, have significantly lower multiplicities in the current hemisphere 
of the Breit frame. It thus appears
that \ep and \e+e- data share a flattening of the KNO distribution at low
energies but that there is an additional 
sensitivity to the effects of higher order QCD processes in the Breit frame 
which do not contribute to \e+e- events.

\section{Summary}

        The Breit frame fragmentation function analysis 
has been repeated with the higher statistics of 1994 data and is
now extended to give multiplicity {\it distributions} as functions of both
$x$ and $Q^2$. Invariant energy spectra are shown to be incompatible with a 
calculation utilising a constant value for $\alpha_s$. We discuss a 
method of discriminating against events in kinematic regions dominated by 
QCD processes which do not contribute to \e+e- interactions. 
We show that some discrepancies between \ep Breit frame current hemisphere 
and \e+e- average charged multiplicities, including an observed violation of 
KNO scaling at low $Q^2$, decrease as conditions are progressively made more 
comparable. With a selection which gives similar kinematic conditions to that 
of a single hemisphere \e+e- annihilation, the \ep energy spectra can be shown 
to be compatible with a common, $Q^2$-independent, low momentum limit. 
In summary, the data 
give strong support to the concept of quark fragmentation universality.

\section*{Acknowledgments}

We are grateful to the HERA machine group whose outstanding
efforts have made and continue to make this experiment possible. We thank
the engineers and technicians for their work in constructing and now
maintaining the H1 detector, our funding agencies for financial support, the
DESY technical staff for continual assistance, and the DESY directorate for the
hospitality which they extend to the non--DESY members of the collaboration.
We are grateful to S.\ Lupia for kind assistance with obtaining MLLA/LPHD
predictions.

%\newpage

\end{document}